# Apparent fractionation of isotopes in moderately cooled argon

## A. R. Cholach[a,*] and D. V. Yakovin[b]

[a]Boreskov Institute of Catalysis, Akademik Lavrentiev Ave 5, 630090 Novosibirsk, Russia

[b]Institute of Automation and Electrometry, Akademik Koptyug Ave 1, 630090 Novosibirsk, Russia

**Abstract**

The fractionation of isotopes of natural Ar near the condensation ($T_c$) and freezing point has been studied using mass spectrometry (MS), numerical modeling and density functional theory. The heat of formation of 0.30, 0.52 and 0.70 kJ per Ar atom of the clusters $Ar_2$, $Ar_3$ and $Ar_4$, respectively, shows the tendency of Ar to clusterization. At $T > T_c$ apparent separation coefficients $\alpha_{40} = 0.80$, $\alpha_{38} = 1.07$ and $\alpha_{36} = 1.28$ for $^{40}Ar$, $^{38}Ar$ and $^{36}Ar$, respectively, are caused by the formation of dimers, which absence during the MS analysis at room temperature indicates their retention in the cryostat. At $T < T_c$, fractionation increases ($\alpha_{40} = 0.56$, $\alpha_{38} = 2.24$ and $\alpha_{36} = 1.67$) due to a difference of 170-520 J/mol between the heat of condensation and dissolution of $^{38}Ar$ and $^{36}Ar$ in $^{40}Ar$ condensate. The formation of a solid phase occurs with preferential freezing out of $^{40}Ar$ ($\alpha_{40} = 0.69$, $\alpha_{38} = 1.57$ and $\alpha_{36} = 1.42$) for kinetic reasons.

*Corresponding author (Email: cholach@catalysis.ru)

## 1. Introduction

Isotope separation is an ongoing area of research, as isotope-enriched compounds are widely used in medicine, chemical technology, materials science, nuclear energy, and many other fields [1-3]. In general, the problem boils down to the separation of components with similar physical and chemical properties. Fractional distillation is not applicable for the separation of components with close boiling points, therefore isotopic enrichment is carried out using special methods based on the difference in properties of the target product and impurities by weight, solubility, adsorption and diffusion ability, etc. Special methods, as a rule, are characterized by high energy intensity, duration and the need for expensive equipment [4-6]. In this regard, the creation of effective and economical methods of isotope separation is an urgent task in the field of chemistry and chemical technology.

The velocity, as well as the de Broglie wavelength, of a light isotope is always higher than that of a heavier one [7]. For this reason, the ratio $P_{36}/P_{40} = 1.0058$ and $1.0021$ of the equilibrium pressures of the isotopes $^{36}$Ar and $^{40}$Ar at $T = 88.2$ K and $118.7$ K, respectively, was found using mass spectroscopic measurements [8]; in a similar study, $P_{36}/P_{40} = 1.0066$ and $1.0061$ at $T = 84.4$ K and $87.0$ K were reported, respectively [9]. It was found that the ratio of equilibrium pressures for the liquid phases $^{36}$Ar and $^{40}$Ar varies from $1.00657$ to $1.00425$ with an increase in temperature from $83.808$ K to $101.418$ K, respectively [10], in accordance with the theoretical model [11]. The heat and temperature of the phase transition of a heavy isotope is higher than that of a light isotope, for the same kinetic reason. This feature underlies the methods of water purification from salts and heavy isotopes [12, 13], since $D_2O$ has a melting point of $3.82$ °C and a boiling point of $101.42$ °C [14, 15].

Dimer formation was detected during supersonic free jet expansion of Ar [16, 17]. The fractions of $0.010$ and $0.014$ dimers in the condensed state were determined at beam temperatures of $47.4$ K and $39.3$ K, respectively, using Rayleigh scattering [16]. The molar fraction of the dimer $0.0007$ was determined at a beam temperature below 10 K using mass spectrometry [17].

Previously, we proposed a method for deep purification of $NF_3$ from $CF_4$ with boiling points of $144.4$ K and $145.2$ K, respectively, at the phase interface under conditions in which only $NF_3$ condenses and the greatest enrichment of the $CF_4$ gas phase is ensured [18]. The purpose of this work is to evaluate the effectiveness of partial condensation and freezing out methods for isotope separation at the gas-liquid and liquid-solid interface of natural argon. The studies were carried out under cryogenic conditions, taking into account the boiling point $T_b = 87.30$ K at $101.325$ kPa and the normal melting point $T_m = 83.81$ K of natural Ar [14].

## 2. Experimental

We used Ar with 99.999% purity (Promgaz LLC, Novosibirsk, Russia) with a natural isotope content. A 0.5-liter stainless steel cylinder was used as a cryostat; the cylinder walls were sanded and wrapped in three layers with 0.5 mm thick copper foil with a free "edge" 10 cm below the bottom. The temperature in the cryostat placed in a stainless-steel thermostat in the range of 78-300 K was set by cooling a part of the copper "edge" with liquid nitrogen. The temperature was measured using a $T$-type thermocouple with an accuracy of 1 K, mechanically attached to the cylinder near the free copper "edge". Pressure measurement in the range of 0.1 – 20 MPa in the cylinder and sampling system was carried out using ELPG300 and Proccontrol HE-200/10bar sensors with an accuracy of 0.5% and 0.2%, respectively. The cylinder valve and the pressure sensor with a total volume of ~ 0.2 liters were at room temperature $T_0$, so the pressure $P$ in the cryostat differed from the measured value $P_{exp}$. The dependence $P(T)$ was calibrated at a reliable experimental cryostat temperature $T_{exp} = T$ as follows. We define $T^*$ as the effective temperature, $T_0 > T^* > T$, providing a molar amount of Ar $n_0 = P_{exp}V_0/RT^*$ in a volume of $V_0 = 0.52$ liters, and $\delta_{hot}$ is the hot part of $V_0$ at $P_{exp}$ and $T_0$; then the cold part is $(1 - \delta_{hot})$ at $P_{exp}$, $T$. The material balance $n_0 = n$ (hot part) $+ n$ (cold part) gives an updated pressure value:

$$P(T) = P_{exp} \left( 1 - \delta_{hot} \left( 1 - \frac{T}{T_0} \right) \right) \quad (1|)$$

where $P_{exp}$ is the experimental pressure, $\delta_{hot}$ is the hot part of the cryostat at $T_0$, $T$ is the reliable experimental temperature.

The isotope composition in the gas samples was determined at room temperature using a monopole mass spectrometer with an accuracy of ~3%, which was estimated from the average deviation of the ion current in 15-20 measurements. The cryogenic system was characterized by total ($t_{tot}$) and local ($t_{loc}$) relaxation times, which correspond to the establishment of equilibrium and equalization of pressure during the phase transition or dimerization reaction, respectively [19]:

$$t_{tot} = \frac{L^2}{2D}; \; t_{loc} = \frac{\lambda^2}{2D} \quad (2)$$

here $D = \frac{1}{3}\bar{v}\lambda$ is the self-diffusion coefficient at the average molecular velocity $\bar{v} = \sqrt{\frac{8RT}{\pi m}}$ and free path length $\lambda = \frac{RT}{\sigma P N_A \sqrt{2}}$ at pressure $P$ and effective cross section $\sigma = \pi d^2$, where the kinetic diameter $d = 3.35$ Å is the average value of experimental measurements $D$ at various pressures and temperatures [20]; $L = 3.5$ cm is the characteristic size of the cryostat; $N_A$ is the Avogadro constant, and $R$ is the gas constant.

The deviation between the experimental and calibrated pressure in Fig. 1a does not exceed 3% at optimal $\delta_{hot} = 0.13$. The holding time in the cryostat at each temperature was 5-10 minutes, which indicates nonequilibrium conditions and incomplete mixing of isotopes in the system (Fig. 1b).

The separation coefficients were determined by the equation $\alpha_i = \frac{x_i/(1-x_i)}{x_{i,0}/(1-x_{i,0})}$, where $x_{i,0} = \frac{n_{i,0}}{\sum n_{i,0}}$ and $x_i = \frac{n_i}{\sum n_i}$ are the initial and final atomic fractions of the $i$-th isotope with a total number of moles of $n_{i,0}$ and $n_i$, respectively [21].

## 3. Theoretical

Density Functional Theory (DFT) calculations with spin-orbital coupling have been performed using the Quantum Espresso package, a nonlocal exchange-correlation functional in the Perdew-Burke-Ernzerhof parameterization with Projected Augmented Wave pseudopotentials and a kinetic cutoff energy of 65 Ry for a plane-wave basis set [22-24]. The Brillouin zone was integrated on a 20 × 20 × 20 Monkhorst-Pack grid using a Gaussian smearing of 0.4 eV [25]. $Ar_n$ clusters ($n = 2$-4) of fully relaxed atoms were modeled using a cubic cell with a lattice parameter of 12.700 Å and a 10 × 10 × 10 mesh. The reaction heats $nAr \rightarrow Ar_n$ were calculated as the difference between the total energies of the reaction products and reagents. Numerical integration of systems of ordinary differential equations corresponding to the formation and dissociation of dimers was carried out by the Euler method using the QBasic software. [26].

## 4. Results and discussion

DFT calculations have shown that the heat of formation ($Q_{cl}$) of small $Ar_n$ clusters ($n = 2$-4) per atom increases with increasing $n$, indicating a tendency for Ar to cluster; the $Q_{cl}$ values in Table 1 are compared with the heat of condensation (evaporation) $Q_{cond} = 6.43$ kJ/mol of natural Ar at $T_b = 87.3$ K [14]. The calculated dimerization heat of 0.5954 kJ/mol is lower than the experimental value of 1.0105 kJ/mol obtained from the vibrational spectra of Ar dimers under collisionless conditions at 40 K [27]. The optimized Ar-Ar distance in small clusters of 3.9934 Å exceeds the Ar-Ar distance of 3.767 Å of the potential minimum "ab initio" of the $Ar_2$ dimer [28].

Table 2 shows the isotope content in the samples taken from the cryostat at different temperatures; the condensation temperature ($T_c$) was obtained from the equivalence of equilibrium pressure and partial pressure $^{40}Ar$ $\bar{P}(T_c) = P_0 x_{0,40} T_c / T_0$. The isotopic composition of the initial Ar is close to the literature data on the natural content of isotopes [14, 29].

The ratio of equilibrium pressures $\frac{\bar{P}(Ar_{36})}{\bar{P}(Ar_{40})} < 1.002$ for kinetic reasons and the corresponding separation coefficient $\alpha < 1.004$ [8, 10] is lower than the essentially same ratio of atomic fractions $x_{36}/x_{36,0} = 1.2796$ and $\alpha_{36} = 1.2807$, respectively. Significant fractionation of isotopes at $T = 152.4$ K above $T_c$ = 134.2 K indicates the following: (1) Dimerization of Ar occurs when cooled to a moderate temperature; (2) Dimers are retained in the cryostat, most likely by condensation [16], since mass spectrometric analysis is performed at room temperature, which excludes the presence of dimers. The behavior of the system in the range of temperatures of the gas phase, condensation and freezing may be different for kinetic reasons [8-10], therefore these areas are considered separately.

## 4.1. Gas phase only

The formation of $^{40}Ar^nAr$ dimers ($n$ = 40, 38, 36) in the gas phase at $T > T_c$, excluding dimers of minor isotopes due to the low content of $^{38}Ar$ and $^{36}Ar$, is described as follows:

$$^{40}Ar + {}^nAr \leftrightarrow {}^{40}Ar^nAr$$

$$\bar{K}_1 = exp\{-\Delta G/RT\} \qquad (k1=3)$$

$$k_f = k_r\bar{K}_1; \; k_r = k_{r,0}exp\{-\Delta H/RT\}$$

where $\bar{K}_1$ is the equilibrium constant, and $k_f$ and $k_r$ are the constants of the forward and reverse reactions, which are considered the same for all isotopes and dimers, respectively.

The Gibbs energy change $\Delta G$ = 5947.70 J/mol and the enthalpy change $\Delta H$ = -3472.74 J/mol in Eq. (3) were found by extrapolating the dependencies $\Delta H(T) = -230.4818 - 21.2746 \cdot T$ and $\Delta G(T) = -2143.0218 + 53.0887 \cdot T$ for dimerization of $^{40}Ar$ [30]. We used a preexponential coefficient $k_{r,0} = \frac{kT}{h}$, close to that $3.85953 \cdot 10^{12}$ s$^{-1}$, corresponding to the vibration frequency $v$ = 128.74 cm$^{-1}$ of the ground state Ar$_2$ at 78 K [31].

The retention of dimers in the cryostat corresponds to the following ratios:

$$Ar_{2,gas} \leftrightarrow Ar_{2,tr}$$

$$\bar{K}_2 = \frac{P(Ar_{2,tr})}{P(Ar_{2,gas})} = exp\{Q_{tr}/RT\} \qquad (k2=4)$$

$$x(Ar_{2,gas}) = \frac{1}{1+\bar{K}_2}; \; x(Ar_{2,tr}) = \frac{\bar{K}_2}{1+\bar{K}_2}$$

where $x(Ar_{2,gas}) = \frac{P(Ar_{2,gas})}{P(Ar_{2,tot})}$ and $x(Ar_{2,tr}) = \frac{P(Ar_{2,tr})}{P(Ar_{2,tot})}$ are the molar fractions of gaseous and trapped Ar$_2$ with effective partial pressures $P(Ar_{2,gas})$ and $P(Ar_{2,tr})$, respectively, and total pressure $P(Ar_{2,tot}) = P(Ar_{2,gas}) + P(Ar_{2,tr})$; $\bar{K}_2$ is the equilibrium constant, $Q_{cond}$ and $2Q_{cond}$ are reasonable lower and upper limits of $Q_{tr}$, respectively; we used $Q_{tr} = Q_{cond}$.

If we ignore the minor isotopes and consider only $^{40}Ar$, then the equilibrium and dynamic reaction modes (3) correspond to the equations:

Equilibrium mode: $P_1 = \sqrt{\left(\frac{P^\emptyset}{4\overline{K_1}}\right)^2 + \frac{C_0 P^\emptyset}{2\overline{K_1}}} - \frac{P^\emptyset}{4\overline{K_1}}$

Dynamic mode: $P_4(t) = \frac{m+n-(m-n)exp\{2n(4k_1 t/P^\emptyset - C_1)\}}{1-exp\{2n(4k_1 t/P^\emptyset - C_1)\}}$

$$(4=5)$$

Both modes: $P_1 + 2P_4 = C_0$

where $m = \frac{C_0}{2} + \frac{P^\emptyset}{8\overline{K_1}}$; $n = \sqrt{m^2 - \frac{C_0^2}{4}}$, $C_0 = P_0 \frac{T}{T_0}$, $C_1 = \frac{1}{2n} ln \frac{(m-n)}{(m+n)}$, and $P^\emptyset = 100$ kPa is the standard pressure.

If we consider all isotopes, the reaction rates (4) correspond to ordinary differential equations:

| Reaction | Equation | |
|---|---|---|
| $^{40}\mathrm{Ar} + {}^{40}\mathrm{Ar} \rightarrow {}^{40}\mathrm{Ar}^{40}\mathrm{Ar}$ | $\dot{P_1} = -P_1 k_f (2P_1 + P_2 + P_3)/P^\emptyset + k_r(2P_4 + P_5 + P_6)$ | |
| $^{40}\mathrm{Ar} + {}^{38}\mathrm{Ar} \rightarrow {}^{40}\mathrm{Ar}^{38}\mathrm{Ar}$ | $\dot{P_2} = -k_f P_1 P_2/P^\emptyset + k_r P_5$ | |
| $^{40}\mathrm{Ar} + {}^{36}\mathrm{Ar} \rightarrow {}^{40}\mathrm{Ar}^{36}\mathrm{Ar}$ | $\dot{P_3} = -k_f P_1 P_3/P^\emptyset + k_r P_6$ | (k3=6) |
| $^{40}\mathrm{Ar}^{40}\mathrm{Ar} \rightarrow {}^{40}\mathrm{Ar} + {}^{40}\mathrm{Ar}$ | $\dot{P_4} = k_f P_1^2/P^\emptyset - k_r P_4$ | |
| $^{40}\mathrm{Ar}^{38}\mathrm{Ar} \rightarrow {}^{40}\mathrm{Ar} + {}^{38}\mathrm{Ar}$ | $\dot{P_5} = k_f P_1 P_2/P^\emptyset - k_r P_5$ | |
| $^{40}\mathrm{Ar}^{36}\mathrm{Ar} \rightarrow {}^{40}\mathrm{Ar} + {}^{36}\mathrm{Ar}$ | $\dot{P_6} = k_f P_1 P_3/P^\emptyset - k_r P_6$ | |

here $k_f$ and $k_r$ are the constants of the forward and reverse reactions, and $P_1 = P(^{40}\mathrm{Ar})$, $P_2 = P(^{38}\mathrm{Ar})$, $P_3 = P(^{36}\mathrm{Ar})$, $P_4 = P(^{40}\mathrm{Ar}_2)$, $P_5 = P(^{40}\mathrm{Ar}^{38}\mathrm{Ar})$, $P_6 = P(^{40}\mathrm{Ar}^{36}\mathrm{Ar})$ are partial pressures.

Numerical integration of the system of equations (6) was performed by the Euler method [26]. This assumes that the current pressure $\{P_i(t_i)\}$ at time $t_i = t_0 + ih$ ($i = 1, 2, …, n$) and under initial conditions $\{P_i(t_0)\} = \{P_0\}$ can be approximated as $P_{i+1} = P_i + hf(P_i, t_i)$, where $h$ is the integration step. The translation of the parameter $h$ in time (1 step = 2.00076 seconds) was performed by comparing the simulated and explicit kinetics of $P_1(t)$ at $k_r = 0$, excluding other steps, which corresponds to the equation $\dot{P_1} = -\frac{2k_f P_1^2}{P^\emptyset}$ and the ratios $\frac{d(1/P_1)}{dh} = \frac{d(1/P_1)}{dt} = \frac{2k_f}{P^\emptyset}$. The parameters of the numerical simulation of the system of equations (6) are given in Table 3. The explicit kinetics of the reaction (Eq. 5) coincides with the one modeled for one isotope.

The total atomic fraction and the corresponding separation coefficient in Fig. 2 correspond to the actual isotope content, while the apparent fraction and separation coefficient take into account the dissociation of dimers during MS measurements at room temperature; then, as follows from Eq. (6):

| Isotope | $x_{tot}$ | $x_{app}$ | |
|---------|-----------|-----------|---|
| $^{40}$Ar | $\dfrac{P_1}{(P_1+P_2+P_3)}$ | $\dfrac{P_1+2P_{4,gas}+P_5+P_6}{P_1+P_2+P_3+2(P_{4,gas}+P_5+P_6)}$ | |
| $^{38}$Ar | $\dfrac{P_2}{(P_1+P_2+P_3)}$ | $\dfrac{P_2+P_5}{P_1+P_2+P_3+2(P_{4,gas}+P_5+P_6)}$ | (k4=7) |
| $^{36}$Ar | $\dfrac{P_3}{(P_1+P_2+P_3)}$ | $\dfrac{P_3+P_6}{P_1+P_2+P_3+2(P_{4,gas}+P_5+P_6)}$ | |

The $P_{4,gas}$ values in Eq. (7) were obtained from Eq. (4), the dimers of minor isotopes were considered uncondensed due to their insignificant content. Dissociation of dimers increases the total number of atoms to a greater extent than $^{40}$Ar isotopes, which leads to a decrease in $x_{40}$ and an increase in the apparent fractionation of isotopes and, thus, to the ratio $\alpha_{app} > \alpha_{tot}$. The modeled molar fractions of $Ar_2$ gas equal to 0.0010, 0.0011 and 0.0016 at 152.4, 126.2 and 83.6 K, respectively, are comparable to the experimental values of 0.0007-0.014 [16, 17], whereas the total molar fractions equal to 0.164669, 0.17317 and 0.250469 are much higher (Equations 4 and 6).

## 4.2. Gas and liquid phases

The transition temperature of the linear dependence $PV = nRT$ to the Clausius-Clapeyron dependence $\bar{P} = P_0 \exp\{\frac{Q_{cond}}{RT}\}$ upon cooling (~131 K) and heating (~132 K) Ar is close to the calculated value (132.0 K) and indicates partial condensation and boiling of the dominant isotope $^{40}$Ar, respectively (Table 2 and Fig. 3).

The dynamics of isotope fractionation above and below $T_c$ is similar, but the experimental separation coefficients in the condensation region exceed the simulated values (Table 4 and Fig. 4). The interaction between condensate atoms is more efficient than between solvent and dissolved atoms, so the heat of condensation of $Q_{cond}$ probably exceeds the heat of dissolution of $Q_{sol}$. The difference $Q_{cond} - Q_{sol}$ ~170-520 J/mol, or ~3-8% $Q_{cond}$, was estimated from the ratio $\frac{x_{i,exp}}{x_{i,sim}} = exp\left\{\frac{Q_{cond}-Q_{sol}}{RT}\right\}$, where $x_{i,exp}$ and $x_{i,sim}$ are experimental and simulated atomic fractions of isotopes.

## 4.3. Solid phase

Table 4 and Fig. 5 show the depletion of a sample taken at a temperature of 83.6 K with an isotope of $^{40}$Ar. Cooling below the freezing point disrupts the formation of dimers, which are most likely to freeze first, and leads to a sequence of separation coefficients $\alpha_{tot} > \alpha_{exp} > \alpha_{app}$.

The experimental pressure of 61.3 kPa, after its sharp drop in Fig. 5f, showing the formation of a solid phase, allows us to estimate the solid fraction of atoms $\delta_s = \frac{n_0-n_g}{n_0} = 0.9622$, where $n_0$ and $n_g$ are the number of moles of the gas phase and the total number of moles, respectively. The separation coefficients in the solid phase $\alpha_{40} = 1.0177$, $\alpha_{38} = 0.9776$ and $\alpha_{36} = 0.9835$, which show its enrichment with $^{40}$Ar, were obtained from the material balance $x_{i,0} = x_{i,s}\delta_s + x_{i,g}(1 - \delta_s)$, where $x_{i,0}$, $x_{i,s}$ and $x_{i,g}$ are the atomic fractions of the $i$-th isotope in the natural Ar, solid phase and gas phase, respectively.

**Conclusion**

The fractionation of isotopes of natural Ar in different temperature ranges relative to the phase transition has been studied using mass spectrometry (MS), numerical modeling and density functional theory (DFT). DFT calculations show the tendency of Ar towards clustering. The separation coefficients $1/\alpha_{40} = 1.25$, $\alpha_{38} = 1.07$ and $\alpha_{36} = 1.28$ for $^{40}$Ar, $^{38}$Ar and $^{36}$Ar, respectively, which were obtained in the gas $T$-region above the boiling point, are higher than expected for kinetic reasons $\alpha < 1.004$. This is due to the formation of dimers, which are retained in the cryostat, probably as a result of condensation, since the MS analysis was performed at room temperature. Fractionation increases in the condensation region ($\alpha_{40} = 0.56$, $\alpha_{38} = 2.24$ and $\alpha_{36} = 1.67$) due to a 3-8% difference between the heat of condensation and dissolution of $^{38}$Ar and $^{36}$Ar in $^{40}$Ar condensate. The enrichment of the $^{40}$Ar solid phase ($\alpha_{40} = 1.02$, $\alpha_{38} = 0.98$ and $\alpha_{36} = 0.98$) in the freezing region of occurs for kinetic reasons.

**Acknowledgements**

This work was supported by the Russian Science Foundation (Project #23-23-00033). The Siberian Supercomputer Center is gratefully acknowledged for providing supercomputer facilities.

**Table 1.** The optimized Ar-Ar distance ($d$) and the heat of formation ($Q_{cl}$) for $Ar_n$ clusters.

| Cluster | $d$ (Å) | $Q_{cl}$ (kJ/mol) | $Q_{cl}/n$ (kJ/mol) | $Q_{cl}/Q_{cond}$ (%) |
|---|---|---|---|---|
| $Ar_2$ | 3.9934 | 0.5954 | 0.2977 | 9.26 |
| $Ar_3$ | 3.9934 | 1.5736 | 0.5245 | 24.47 |
| $Ar_4$ square | 4.0000 | 2.3709 | 0.5927 | 36.87 |
| $Ar_4$ tetrahedron | 3.9934 | 2.8165 | 0.7041 | 43.80 |

**Table 2.** Sample temperature ($T$), condensation temperature ($T_c$), total pressure ($P_0$) at initial temperature ($T_0$), the ratios of atomic fractions ($x_i/x_{i,0}$) and separation coefficients ($\alpha_i$) of Ar isotopes, as well as the initial content, natural abundance of Ar isotopes and their deviation ($\Delta$).

| $T$ (K) | 83.6 | 126.2 | 152.4 | This work | [29] | $\Delta$ (%) |
|---|---|---|---|---|---|---|
| $T_c$ (K) | 132.0 | 130.2 | 134.2 | - | - | - |
| $P_0$ (MPa) | 5.52 | 5.15 | 5.95 | 5.95 | - | - |
| $T_0$ (K) | 300.2 | 301.1 | 298.2 | 300 | - | - |
| $x_{36}/x_{36,0}$ | 1.4179 | 1.6625 | 1.2796 | - | - | - |
| $x_{38}/x_{38,0}$ | 1.5685 | 2.3278 | 1.0701 | - | - | - |
| $x_{40}/x_{40,0}$ | 0.9982 | 0.9969 | 0.9990 | - | - | - |
| $\alpha_{36}$ | 1.4200 | 1.6662 | 1.2807 | 0.3391 | 0.3365 | 0.767 |
| $\alpha_{38}$ | 1.5693 | 2.3297 | 1.0701 | 0.0628 | 0.0632 | -0.637 |
| $\alpha_{40}$ | 0.6925 | 0.5644 | 0.8013 | 99.5981 | 99.6003 | -0.002 |

**Table 3.** Parameters used in the numerical modeling; $P_{4,0} = P_{5,0} = P_{6,0} = 0$.

| $T$ (K) | $\overline{K}_1$ | $k_r$ (s$^{-1}$) | $P_{1,0}$ (kPa) | $P_{2,0}$ (kPa) | $P_{3,0}$ (kPa) |
|---|---|---|---|---|---|
| 152.4 | 0.009151360 | 2.049120293E+11 | 3027.1570 | 1.90942452 | 10.3053547 |
| 126.2 | 0.013000881 | 1.633961695E+11 | 2304.3809 | 1.453523 | 7.844807 |
| 83.6 | 0.036808983 | 9.677836797E+10 | 1530.49624 | 0.96538338 | 5.21027045 |

**Table 4**. The actual ($\alpha_{tot}$) and apparent ($\alpha_{app}$) separation coefficients of Ar isotopes, obtained by simulation, explicit solution (only for $^{40}$Ar) and experiment ($\alpha_{exp}$).

| $T$ (K) | Isotope | Simulation | | Explicit | | Experiment |
|---------|---------|------------------|------------------|------------------|------------------|-------------------|
| | | $\alpha_{tot}$ | $\alpha_{app}$ | $\alpha_{tot}$ | $\alpha_{app}$ | $\alpha_{exp}$ |
| 152.4 | $^{36}$Ar | 1.1661 | 1.3923 | - | - | 1.2807 |
| | $^{38}$Ar | 1.1656 | 1.3908 | - | - | 1.0701 |
| | $^{40}$Ar | 0.8575 | 0.7180 | 0.7153 | 0.7171 | 0.8013 |
| 126.2 | $^{36}$Ar | 1.1747 | 1.4168 | - | - | 1.6662 |
| | $^{38}$Ar | 1.1741 | 1.4152 | - | - | 2.3297 |
| | $^{40}$Ar | 0.8512 | 0.7056 | 0.7028 | 0.7035 | 0.5644 |
| 83.6 | $^{36}$Ar | 1.2528 | 1.6641 | - | - | 1.4200 |
| | $^{38}$Ar | 1.2519 | 1.6610 | - | - | 1.5693 |
| | $^{40}$Ar | 0.7981 | 0.6007 | 0.5969 | 0.5970 | 0.6925 |

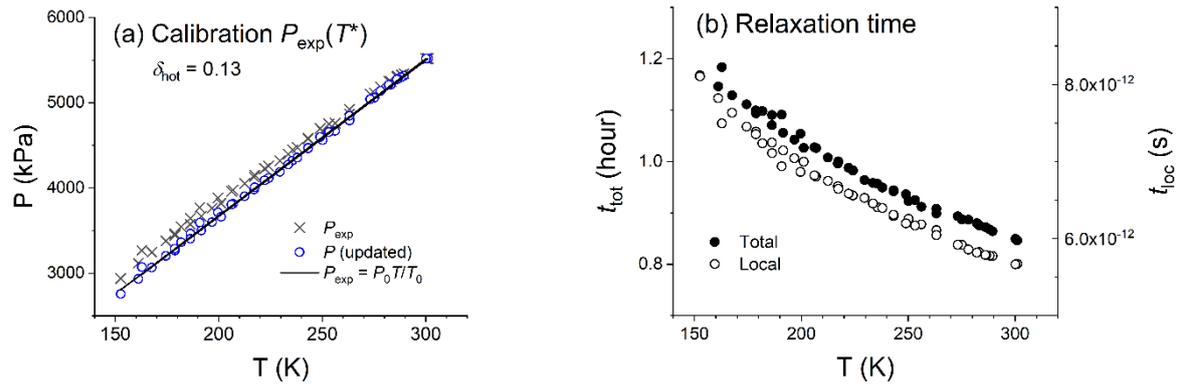

**Fig. 1.** (a) Experimental and calibrated dependences of $P(T)$ (Eq. 1); (b) total and local relaxation time (Eq. 2).

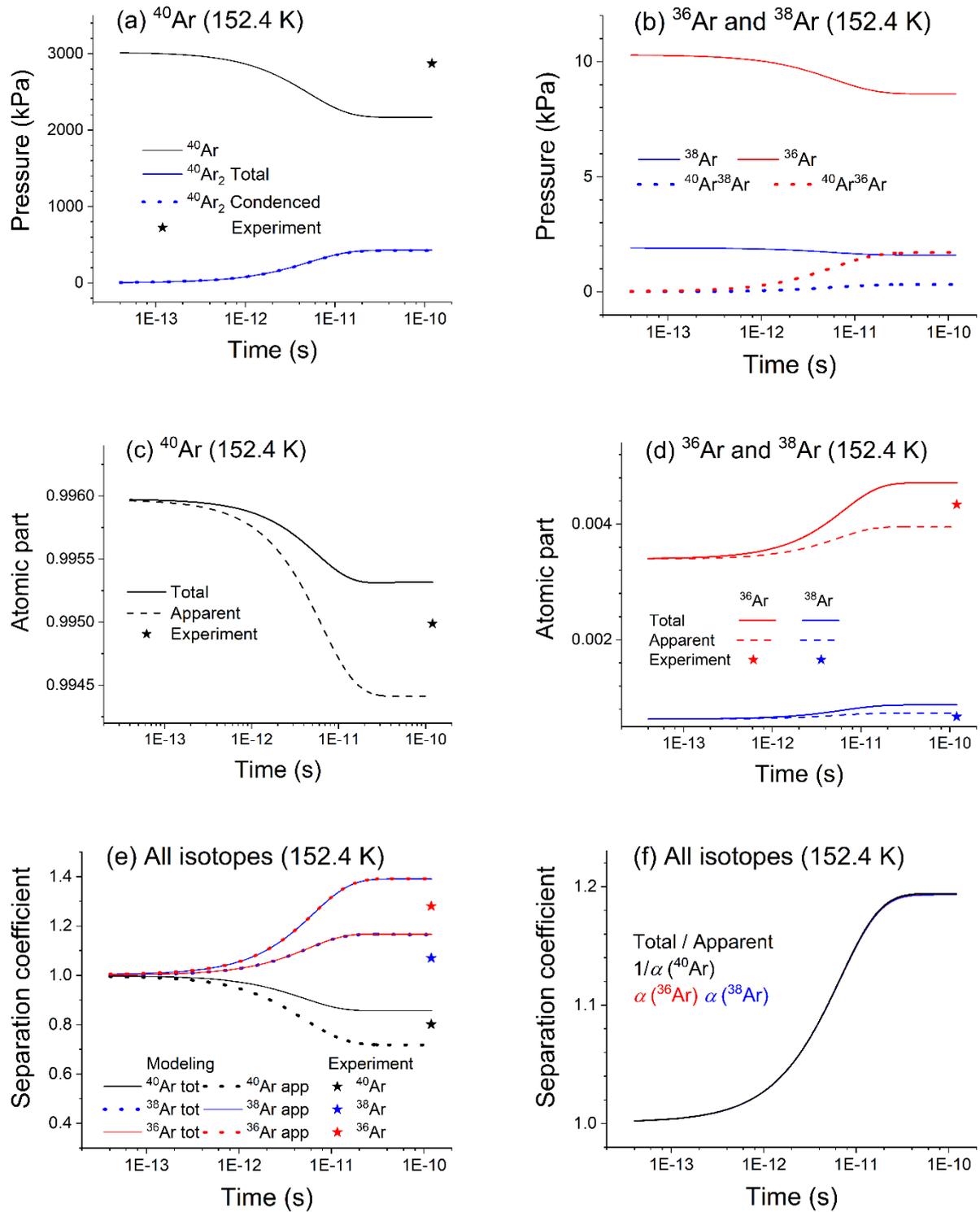

**Fig. 2.** Simulated fractionation kinetics of (a, c) major and (b, d) minor isotopes, (e) separation coefficients and (f) $\alpha_{tot}/\alpha_{app}$ ratio at $T = 152.4$ K in comparison with the experiment (Eq. 6 and Eq. 7).

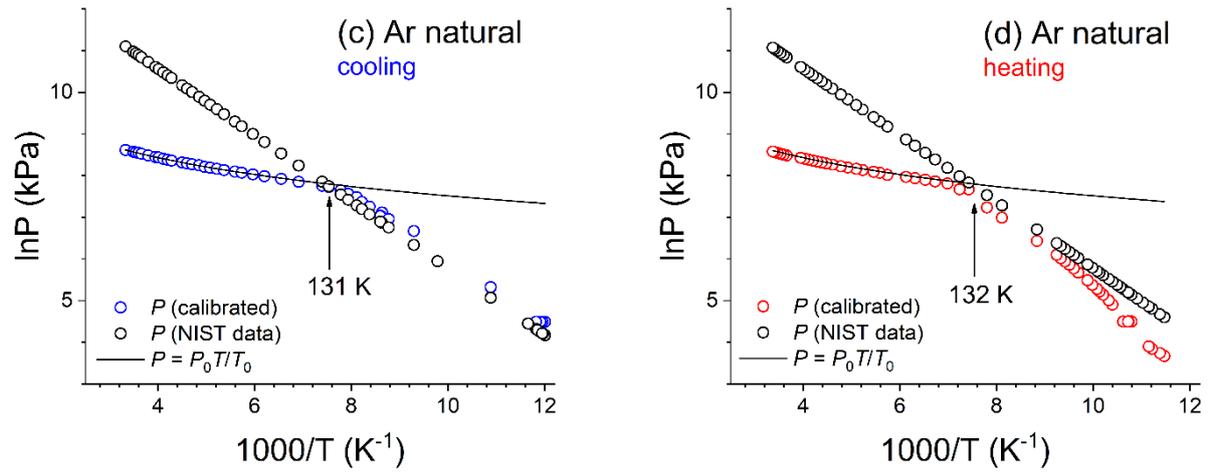

**Fig. 3**. Experimental and calibrated dependence of $P(T)$ on (a) heating and (b) cooling of natural Ar at $P_0$ = 5.52 MPa.

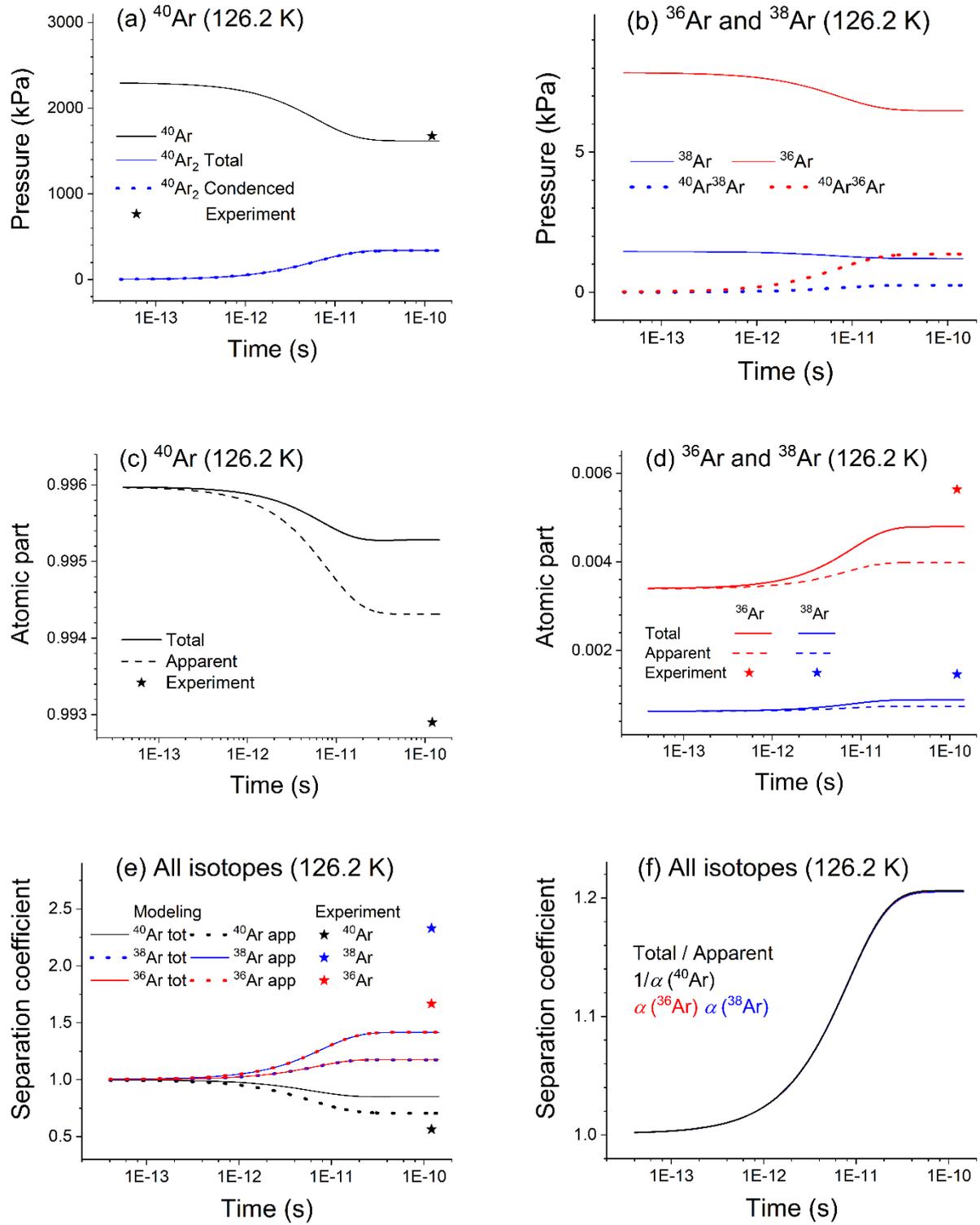

**Fig. 4.** Simulated fractionation kinetics of (a, c) major and (b, d) minor isotopes, separation coefficients (e) and $\alpha_{tot}/\alpha_{app}$ ratio (f) at $T = 126.2$ K in comparison with the experiment (Eq. 6 and Eq. 7).

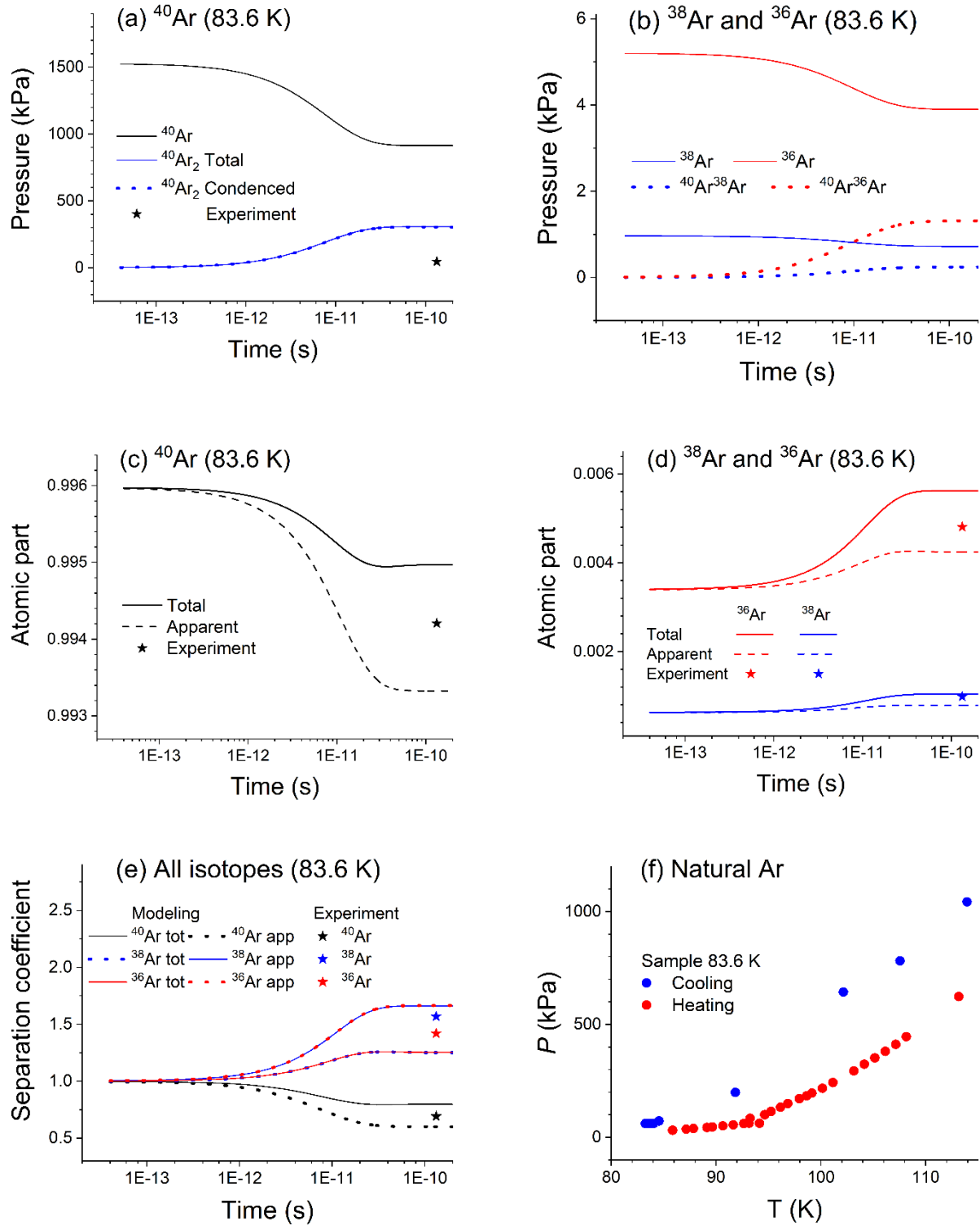

**Fig. 5.** Simulated fractionation kinetics of (a, c) major and (b, d) minor isotopes, separation coefficients (e) and $\alpha_{tot}/\alpha_{app}$ ratio (f) at $T = 83.6$ K in comparison with the experiment (Eq. 6 and Eq. 7).